\documentstyle[preprint,aps]{revtex}
\begin{document}

\title{Positronic lithium, an electronically stable Li--$e^+$ ground 
state}

\author{G.G. Ryzhikh\cite{my} and J. Mitroy}
\address{Faculty of Science, Northern Territory University,
Darwin NT 0909, Australia}

\maketitle

\vspace{1cm}

\begin{abstract}
Calculations of the positron-Li system were performed using the
Stochastic Variational Method and yielded a minimum energy of
$-7.53208$ Hartree for the $L = 0$ ground state.
Unlike previous calculations of this system, the system
was found to be stable against dissociation into the
Ps + Li$^+$ channel with a binding energy of $0.00217$
Hartree and is therefore electronically stable.
This is the first instance of a rigorous calculation
predicting that it is possible to combine a positron with a
neutral atom and form an electronically stable bound state.
\end{abstract}

One of the most tantalizing questions of positron physics is 
whether it is possible for a positron to bind itself to a neutral 
atom and form an electronically stable state \cite{schrad1,schrad2}.
This is a question which can only be answered by a sophisticated
calculation (or experimentation) as the mechanisms responsible
for binding the positron to the atom are polarization potentials
present in the positron-atom complex.  The accurate computation of
the polarization potential for a positron--atom (or electron--atom)
system is of course a challenging exercise in many body physics.

While the question of whether it is possible to bind a positron
to a neutral atom is an open question, the ability of positronium
to attach itself to atoms has been known for a long time.  A 
number of previous works have demonstrated that the
positronium--hydride (PsH) species 
\cite{ore,lebeda,clary,ho,frolov,schrad3}
is stable against dissociation into the Ps + H
or the e$^+$ + H$^-$ channels.  In this case, binding is more
likely since the positron is binding itself to a species with an 
overall negative charge. 

The question of whether a positron can form an electronically
stable bound state with a neutral atom is more vexing.  Dzuba
et al \cite{dzuba} have made calculations suggesting that it is
possible
to bind a positron to atomic species with two valence electrons such
as Mg, Zn, Cd and Hg.  These calculations were performed in the
framework of many--body perturbation theory and their results, while
suggestive, cannot be regarded as providing proof to the
existence to electronically stable positronic atoms.

In their work, Dzuba et al \cite{dzuba} did not consider the
possibility of positrons forming bound states with alkali atoms
such as Li, Na, K, ... even though the polarization potential
for these species should be stronger than for the alkaline
and alkaline earth atoms and therefore the possibility
of binding should be improved.  One difficulty in binding
positrons to alkali atoms is that the ionization energy
of the alkali atoms is smaller than the binding energy of
positronium.   Therefore the binding energy of the positron
to the neutral atom must exceed a particular value for the
species to be stable against dissociation into positronium + ion.
For example, for the Li--e$^+$ species to be stable against
dissociation into positronium plus Li$^+$ requires that the binding
energy of the e$^+$ with respect to the Li ground state be greater
than $(0.25 - 0.19815)$ Hartree.  In this respect, it is more 
appropriate to regard the possibility of binding as a question 
of whether Ps can itself bind to Li$^+$.

Previous works \cite{clary,japan} on this species had shown that
while the Li--e$^+$ system can have a total energy lower than 
neutral Li, the energy was not low enough to prevent dissociation
into Ps + Li$^+$.  In this work, a large variational calculation
of the Li--e$^+$ system is performed using the Stochastic Variational
Method of Varga and Suzuki \cite{varga2}.  In contradiction with 
previous works it is found that the ground state is electronically 
stable and the binding energy for this state is calculated.

A Gaussian basis has long been a
popular tool for variational calculations in various areas of
quantum physics and chemistry. The Gaussian basis used in this
work has two very important features that make it possible to
generate very accurate wave functions for few body systems.
First, the matrix elements of the interaction Hamiltonian can 
be calculated analytically or, at worst, reduced to a
one-dimensional integral for any number of particles.  Second, that
part of the wave function concerned with the spatial coordinates
maintains its functional form after any possible permutation of
the particles. This is a very useful property for studying
systems containing identical particles.

The Stochastic Variational Method (SVM) was initially proposed
as a method suitable for solving nuclear structure problems
involving a small number of particles \cite{kuku1,kuku2}.
The main idea behind the method is to use stochastic
techniques to optimize the non-linear parameters (i.e. the
exponents) of the underlying Gaussian basis. Since the Gaussian
basis contains terms with $r_{ij}^2$ correlation factors, the
method is capable of achieving results of the highest accuracy
provided the non-linear parameters are properly optimized.

In recent years, the SVM and related methods have been used by
many groups to perform high precision variational calculations in
atomic, mesoatomic, hypernuclear and multiquark systems
\cite{frolov,kopyl,koles,barb}.  Recently, the SVM has been
modified to allow the calculation of excited states and also 
to permit the use of a wide variety of non-central forces 
\cite{varga1,suzuki1}.  In this work, the program of Varga and 
Suzuki \cite{varga2} (which can be used with arbitrary pairwise
central forces) was used for the calculations. A detailed
description of the method and the results of test calculations
on various atomic and nuclear system containing 3--6 particles
can be found in \cite{varga2,varga1}.

An initial series of calculations on a variety of related 
species were performed to estimate the uncertainties in the 
present calculation and validate the method. Results of our 
calculations for neutral Li, neutral Be, and the PsH species 
are shown in Table \ref{tab1} and compared with other accurate 
nonrelativistic calculations.  We show results that were computed 
with an infinite nuclear mass to simplify comparison with the
other results in Table \ref{tab1}.  Our calculation for PsH, agreed 
with the best previous calculation to within 4$\times 10^{-6}$
Hartree \cite{frolov}. Results for the more complicated Be
and Li$^-$ species underestimate the best calculations
\cite{fischer} by less than $7.0\times 10^{-4}$\ Hartree.

Since the question of whether an electronically stable bound state
exists depends on the energy relative to the sum of the energies
for the Li$^+$ and Ps atoms, the energy of the Li$^+$ ground state
was computed.  Our result is identical with that of the classic
calculation of Pekeris \cite{pekeris} to 8 significant
figures and indicates that binding will occur if the total energy
of the Li-e$^+$ system is lower than -7.529913 Hartree.

The convergence of energy of the Li--e$^+$ system as a function 
of the number of gaussoid basis functions is shown
in the Table \ref{tab2}.  It is noticeable that a very
large calculation, including at least 300 gaussoid basis
functions, was needed before definite evidence of
a bound state was obtained.  The largest calculation 
included 800 basis functions, and resulted in a total 
energy of $E = -7.53208$ Hartree which is equivalent to 
a binding energy of $\varepsilon = 0.00217$ Hartree.  
When reference is made to the binding energy of
the Li--e$^+$ system it should be noted that the binding
energy is relative to breakup into Li$^+$ and Ps.

Energy expectation values were also computed with the present
optimized wave functions for a finite mass.  The $^7$Li
nucleus has a mass of $M=12863.2 m{_e}$ and for this
species we obtained $E(300) = -7.279325$ Hartree for Li$^+$,
and $E(800) = -7.531491$ Hartree for Li--e$^+$ giving
a binding energy of $\varepsilon = 0.00217$ Hartree.  For
most purposes, finite mass effects can be ignored
since they will not change the binding energy by more than 1\%.

The statement that a bound state exists also remains
valid when relativistic effects are taken into consideration.
One estimate of the relativistic energy correction for neutral
Li is 0.000011 Hartree \cite{davidson}.  An energy correction of
this size cannot affect the primary conclusion, namely the existence
of an Li--e$^+$ bound state, but might have to be taken into
consideration if a really precise value of the binding energy
is to be achieved.  Nevertheless, we are confident in asserting
that the Li--e$^+$ ground state is electronically stable against
decay into both the Li--e$^+$ and Li$^+$--Ps channels.

While the state is electronically stable, it is not stable against
electron-positron annihilation.  The dominant decay process for
electron-positron annihilation is into two $\gamma$--rays.
Therefore the two-photon annihilation rate $\Gamma_{2\gamma}$
was computed using the general formula,
\begin{equation}
\Gamma_{2\gamma} = \pi n \alpha^4 c a_0^{-1} <\delta_{-+}>\ \approx \
50.30874045 \times 10^9\ n <\delta_{-+}> {\rm sec}^{-1},
\end{equation}
which is valid for a system containing $n$ electrons and one
positron \cite{frolov}.  In the above expression,
$\delta_{-+}$ is the expectation value of electon-positron Dirac
$\delta$ function
\begin{equation}
<\delta_{-+}> = \frac{<\Psi|\delta({\bf r}_{e^-} - {\bf 
r}_{e^+})|\Psi>}
{<\Psi|\Psi>}
\end{equation}
The annihilation rate for the Li--e$^+$ system was
$\Gamma_{2\gamma} = 1.70\times 10^9\  {\rm sec}^{-1}$.  
The annihilation rate for PsH has been computed 
as a consistency check and the value we obtain,
$\Gamma_{2\gamma} = 2.45\times 10^9\  {\rm sec}^{-1}$,
is consistent with the best previous estimate \cite{frolov},
namely $\Gamma_{2\gamma} = 2.436\times 10^9\ {\rm sec}^{-1}$.

Other recent studies of the positron--Li system \cite{clary,japan}
had shown that an electronically stable bound state
did not exist.  The failure to find a bound state can be attributed
to the difficulty in performing a calculation on a system containing
4 active particles that had to be accurate to 10$^{-3}$ Hartree.
The most recent study \cite{japan} of the Li--e$^+$ system used the
Diffusion quantum Monte Carlo method to predict an energy of
$-7.5203\pm 0.0048$ Hartree which only just failed to indicate a
stable bound state.  This calculation, correctly predicted the binding
energy of the e$^+$--H$^-$ system ($-0.7891\pm 0.0020$ Hartree) but
evidently the calculation of Li--e$^+$ system is more exacting.

The configuration-interaction-Hylleraas calculation (CI-Hy) of 
Li--e$^+$ system by Clary \cite{clary} was performed by adapting 
a method that had previously
been very successful for atoms \cite{sims} and gave an energy of
$-7.5094$ Hartree.  As a similar calculation by Clary \cite{clary}
of the PsH system underestimated the energy by $0.0050$
Hartree it is not unexpected that it failed to predict a stable
Li--e$^+$ system.  Given that the CI-Hy method 
\cite{sims} gave an energy for neutral Be (-14.6665 Hartree)
which agrees with the best current estimate to within 0.0008
Hartree it is interesting to speculate on the reason for the
slower convergence of the method for systems containing 
a positron. The resolution of this puzzle probably lies in
the fact that the correlations between an electron and a
positron are distinctly different than the correlations between
two electrons.  A system involving purely electrons has two implicit
features that will act to diminish the importance of inter-electronic
correlations.  First of all, the Pauli principle acts to
keep electrons with the same spin away from each other.
Second, the electron-electron interaction also acts to keep
electrons away from each other.  However, neither of these
effects is present if an electron is replaced by a positron.
The interaction between an electron and a positron is attractive,
and it easy to imagine a system with one valence electron 
like lithium evolving into a configuration consisting of a 
positronium atom orbiting around a positively charged 
$(1s)^2$ core.

This possibility was investigated
by projecting the Li--e$^+$ ground state wave function onto
a wave function containing the product of the ground state
positronium wave function and the two electron wave function
for Li$^+$.  The normalization of the residual part of the
projected wave function (essentially the wave function for
the Ps center of mass) was found to be 0.93.  Therefore, 
the best heuristic model of the Li--e$^+$ ground
state would be to regard the system as a positronium atom
weakly attached to, and orbiting around a Li$^+$ $(1s)^2$
core.

The present calculation represents the first rigorous calculation
giving positive evidence that it is possible to combine a positron
with a neutral atom and form an electronically stable system.
Although, the best ab-initio estimate of the binding energy, 
0.00217 Hartree is subject to uncertainties due to incomplete 
convergence of the Li--e$^+$ energy, the statement that the system 
is electronically stable will certainly remain valid under any 
possible refinements of the model.

Having shown that it is possible to combine a positron with
neutral Li to form an electronically stable bound state, an
immediate question arises as to whether it is possible to
join a positron to a more complicated alkali atom such as
sodium and also form a bound state.  The answer to this
question cannot be obtained with a calculation identical to
the present calculation, rather the present method would
have to be refined to incorporate the physics of a closed
shell core.  The possible existence of additional positronic
atoms is a topic that is worth further investigation.

The authors would like to thank K. Varga for the use of his
SVM program and for useful discussions.

\vspace{1cm}

\begin{table}[t]
\caption[]{Non-relativistic energies (in Hartree) of
various atomic systems compared with previous accurate results.
In these calculations the nuclear mass has been assumed to
be infinite. The number in parentheses refers to the total
dimension of the gaussoid basis.}
\label{tab1}

\vspace{8mm}

\begin{tabular}{ccc}
System            & $E$ (SVM, this work) & $E$
("best" nonrelativistic) \\ \hline
Li$^+$            & --7.2799133 (300) & --7.2799133$^a$ \\
PsH               & --0.789183 (400) & --0.789179 $^b$ \\
Li                & --7.478041 (400) & --7.4780603$^c$ \\
Be                &--14.66676  (601) & --14.66732 $^d$ \\
Li + e$^-$        & --7.50012  (600) & --7.50076 $^d$ \\
Li + e$^+$        & --7.53208  (800) & --7.5203  $^e$ \\
\end{tabular}

$^a$ reference \cite{pekeris} \\
$^b$ reference \cite{frolov} \\
$^c$ reference \cite{drake} \\
$^d$ reference \cite{fischer} \\
$^e$ reference \cite{japan}
\end{table}

\vspace{1cm}

\begin{table}[ht]
\caption[]{Convergence of the Li--e$^+$ energy (in Hartree)
as a function of basis size.  Last column shows the energy relative
to the Li$^+$--Ps threshold at -7.529913 Hartree.}
\label{tab2}

\vspace{8mm}

\begin{tabular}{ccc}
$E$ & basis size & $\varepsilon$  \\
\hline
--7.52360           & 200        &  not bound \\
--7.52773           & 300        &  not bound \\
--7.52897           & 350        &  not bound \\
--7.53002           & 400        &  0.00011  \\
--7.53084           & 500        &  0.00093  \\
--7.53135           & 600        &  0.00144  \\
--7.53165           & 700        &  0.00174  \\
--7.53208           & 800        &  0.00217  \\
\end{tabular}
\end{table}

\end{document}